\documentclass[11pt]{article}

\usepackage[T1]{fontenc}
\usepackage[utf8]{inputenc}
\usepackage{lmodern}
\usepackage[letterpaper,margin=1in]{geometry}
\usepackage[authoryear,round]{natbib}
\usepackage{amsmath,amssymb,mathtools}
\usepackage{booktabs,longtable,tabularx,array}
\usepackage{pdflscape}
\usepackage{graphicx}
\usepackage{float}
\usepackage{microtype}
\usepackage{enumitem}
\usepackage{xcolor}
\usepackage{tikz}
\usetikzlibrary{arrows.meta,positioning,fit,backgrounds,calc,decorations.pathreplacing}
\usepackage{url}
\usepackage{hyperref}
\hypersetup{colorlinks=true,linkcolor=black,citecolor=black,urlcolor=black}

\newcommand{\rating}{X_i}

\newcommand{\state}{S_{it}}

\newcommand{\experience}{R_i}

\begin{document}

\title{From Risk Prediction to Risk Mechanisms:\\
A Multi-Resolution Causal Representation for Road Safety and Motor Insurance}
\author{Arthur Charpentier\\[0.5em]
\small Universit\'e du Qu\'ebec \`a Montr\'eal, Montr\'eal, Canada\\
\small Kyoto University, Kyoto, Japan}
\date{August 2026}
\maketitle

\begin{abstract}
Operational risk models can estimate event frequencies precisely while leaving the underlying mechanisms weakly resolved. We study this resolution mismatch using motor insurance and road safety. A revisable DAG represents trip-level crash generation; a separate predictive layer links annual rating information to latent driving states; and an observation process maps crashes into recorded liability claims. Compatibility sets collect the structural and crash-to-claim laws that reproduce an observed annual contrast under stated restrictions. External studies enter only through explicit bridge assumptions and sensitivity bounds. The framework therefore distinguishes sampling uncertainty from uncertainty about structure, observation, and study-to-target correspondence. Two limited examples illustrate the gap. A sublinear mileage relation constrains an aggregate accident rate per unit distance but not its mechanism. In French motor-liability data, the 18--20 versus 40--49 claim-frequency relativity is 3.388 in a model including vehicle and geographic variables and 1.235 when the same model also conditions on a medium-resolution bonus--malus score. These are different predictive functionals, not successive causal adjustments. A Spanish culpability estimate is used only to show how a strong cross-study bridge would restrict a toy bookkeeping region. The framework does not estimate the full DAG; it makes explicit which assumptions are needed before an annual predictive contrast can support a mechanism-specific risk statement.
\end{abstract}


\section{INTRODUCTION}
\label{sec:intro}

Risk models often predict outcomes at a different resolution from the mechanisms that generate them. This matters when the variables used for prediction are aggregated, administrative, or recorded after the physical event. In such settings, statistical precision about a predictive contrast says little by itself about the causal explanation of that contrast. This distinction is familiar in risk analysis, where model form and epistemic uncertainty matter alongside sampling variation \citep{kaplan1981risk,patecornell1996uncertainty,morganhenrion1990,north2020riskcausal}.

Motor insurance provides a clear example. Insurers observe policy characteristics and liability claims over months or years; road-safety research studies road, traffic, vehicle and behavioural states that change within trips. Relating an annual insurance contrast to a crash mechanism therefore requires a bridge across time scales and a second bridge from crash occurrence to the recorded claim.

Let $N_i$ be the annual number of recorded motor-liability claims for policyholder $i$, and let $X_i$ contain rating information such as age, vehicle characteristics, territory, declared use, mileage and prior insurance history. The annual predictive target is
\begin{equation}
  \lambda_{\mathrm{claim}}(X_i)=\mathbb{E}(N_i\mid X_i).
  \label{eq:glm}
\end{equation}
A Poisson GLM is one possible estimator of this mean. The estimand itself, however, is annual claim frequency. It does not describe the sequence of circumstances preceding a collision, and a claim is not the same event as a crash: reporting, coverage, responsibility attribution and claim administration occur downstream.

We use this setting to do three things. We formalize the \emph{resolution mismatch} between annual predictive risk and short-horizon crash mechanisms; separate the crash-occurrence structure from the predictive rating layer and the crash-to-claim process; and define compatibility sets of structural and observation laws that reproduce an annual contrast under stated restrictions. The aim is to show where mechanism-specific interpretation requires assumptions beyond the fitted risk model.

This set-valued perspective draws on partial identification \citep{manski2003partial,tamer2010partial}. Formal transportability asks when causal quantities can be recovered across populations under explicit invariance assumptions \citep{bareinboim2013transport,pearlbareinboim2014external}; when such a transport formula is unavailable, we use weaker sensitivity bounds. Related work on regression sensitivity addresses unmeasured confounding within a fixed estimand \citep{cinelli2020sensitivity,vanderweele2017evalue}. Actuarial models have long distinguished accidents from reported claims and used experience rating to summarize past claim information \citep{boucher2009accidents,boucher2022bonusmalus}. Our focus is the combination of these issues when prediction and mechanism are observed at different resolutions.

Let $H_i$ denote persistent driver heterogeneity, $R_i$ accumulated driving experience, $C_{it}$ the context of driving opportunity $t$, $S_{it}$ a transient driver state, $B_{it}$ immediate behaviour, $K_{it}$ a critical conflict/avoidance state, and $Y_{it}$ the indicator of a crash during that opportunity. Write $V_i=(V_i^{\mathrm{perf}},V_i^{\mathrm{safe}})$ for vehicle-performance and safety mechanisms, and define
\[
 \Omega_{it}=(H_i,R_i,C_{it},S_{it},B_{it},K_{it},V_i^{\mathrm{safe}}),
 \qquad
 p_{it}=\Pr(Y_{it}=1\mid\Omega_{it},X_i,M_i\ge t).
\]
A driving opportunity is a pre-specified, non-overlapping trip or trip segment to which at most one occurrence indicator is assigned. Opportunities are ordered in time; $M_i$ may be random and history-dependent. If $D_{it}$ is distance travelled in opportunity $t$, then $m_i=\sum_{t=1}^{M_i}D_{it}$. Under finite-integrability conditions,
\begin{equation}
\lambda_{\mathrm{crash}}(X_i)
=
\mathbb{E}\!\left[\sum_{t=1}^{M_i}p_{it}\,\middle|\,X_i\right].
\label{eq:central}
\end{equation}
Section~\ref{sec:integrated} defines the annual latent history $U_i$ and its law $Q_x=\mathcal L(M_i,U_i\mid X_i=x)$. The baseline graph uses the strong reduction $p_{it}=\pi_Y(K_{it})$, but the annual identity above does not require it; Section~\ref{sec:graphsensitivity} treats direct paths to $Y_{it}$ as structural sensitivity.

The numerical examples are intentionally narrow. Mileage illustrates an aggregate exposure identity; the age example shows how an endogenous history variable changes a conditional claim contrast. Neither example identifies a causal share of the observed relativity.

\section{RISK PREDICTION AND MECHANISM AT DIFFERENT SCALES}
\label{sec:timescales}

\subsection{Annual predictive risk as a reduced-form target}

Claim-frequency models aggregate heterogeneous driving over a policy year. Separating mileage $m_i$ from the other rating variables, write $X_i=(m_i,Z_i)$. A specification such as
\begin{equation}
  \log \lambda_{\mathrm{claim},i} = \log m_i + \beta^\top Z_i
\end{equation}
treats mileage as exposure and assigns the remaining claim rate per unit distance to variables measured at policy inception or updated infrequently. That representation is useful for pricing and classification, but it leaves the composition of exposure implicit.

Telematics records time of day, speed, braking, acceleration and route-related signals that conventional rating variables leave latent \cite{verbelen2018telematics,baecke2017telematics,henckaerts2022dynamic}. Hidden-state models likewise summarize trip-level behaviour and relate those states to insurance losses \cite{jiang2024hmm}. This evidence supports treating $X_i$ as information about a distribution of driving states, not as a list of direct crash causes.

\subsection{Short-term crash generation}

Evidence closer to collision time covers speed \cite{kloeden1997speed,elvik2019speed}, traffic state \cite{golob2004traffic,xu2012trafficstate,rosandel2015traffic}, weather \cite{qiu2008weather}, sleepiness \cite{martiniuk2013sleep,moradi2019sleepiness}, distraction \cite{simmons2016distraction}, young-driver passenger exposure and behaviour \cite{simonsmorton2005passengers,goodwin2012passengers,ouimet2015passengers}, and novice nighttime or passenger restrictions \cite{fell2011gdl}. Conflict models provide an intermediate scale between ordinary driving and rare crashes \cite{davis2011conflicts}. These studies define the mechanisms represented in the short-term graph; annual rating labels remain in the observation layer.

\begin{figure}[t]
\centering
\begin{tikzpicture}[x=1.2cm,y=1cm,>=Latex]
  \draw[-{Latex[length=2mm]},thick] (0,0) -- (10.5,0);
  \foreach \x/\lab in {0/second,2/minute,4/trip,7/month,10/year}{
    \draw (\x,0.12)--(\x,-0.12);
    \node[below=5pt,font=\small] at (\x,0) {\lab};
  }
  \node[align=center,above=10pt,font=\small] at (1,0) {reaction\\speed\\conflict};
  \node[align=center,above=10pt,font=\small] at (3.2,0) {traffic\\fatigue\\distraction};
  \node[align=center,above=10pt,font=\small] at (5.2,0) {road\\weather\\passengers};
  \node[align=center,above=10pt,font=\small] at (8.7,0) {rating factors\\claim count};
  \draw[decorate,decoration={brace,amplitude=5pt},yshift=34pt] (0,.75)--(6,.75)
     node[midway,yshift=13pt,font=\small]{road safety research};
  \draw[decorate,decoration={brace,amplitude=5pt},yshift=34pt] (7,0.25)--(10.5,0.25)
     node[midway,yshift=13pt,font=\small]{insurance};
\end{tikzpicture}
\caption{The time-scale mismatch. Road safety mechanisms are predominantly studied over seconds, minutes and trips, whereas traditional motor-insurance risk classification aggregates experience over months or years.}
\label{fig:timescale}
\end{figure}
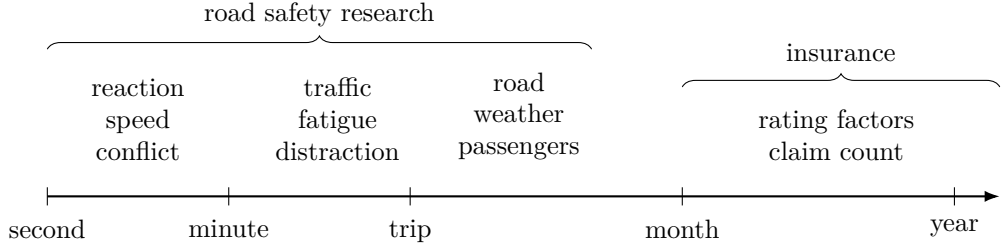

\subsection{From crashes to recorded liability claims}
\label{sec:claimobservation}

The insurance outcome adds an observation process after crash occurrence. Let $J_{it}\in\mathbb N_0$ denote the number of recorded liability-claim contributions attributed, under the portfolio's accounting convention, to opportunity $t$, with $J_{it}=0$ when $Y_{it}=0$. The count-valued definition permits zero, one or more recorded claim contributions from an underlying crash event. Define
\begin{equation}
 \rho_{it}
 =\mathbb E(J_{it}\mid Y_{it}=1,\Omega_{it},X_i,M_i\ge t).
 \label{eq:rho}
\end{equation}
Thus $\rho_{it}$ is a conditional mean claim contribution per crash, not a probability unless the accounting rule makes $J_{it}$ binary. It integrates over post-crash features not represented in the occurrence DAG, including severity, third-party involvement, coverage, reporting, responsibility attribution and claim administration. With $N_i=\sum_{t=1}^{M_i}J_{it}$ under the portfolio's accounting convention, the same finite-integrability conditions used for Eq.~\eqref{eq:central} and iterated expectation give
\begin{equation}
 \lambda_{\mathrm{claim}}(X_i)
 =\mathbb E\!\left[
   \sum_{t=1}^{M_i}p_{it}\rho_{it}
   \,\middle|\,X_i
 \right].
 \label{eq:claimobservation}
\end{equation}
Equality of $\lambda_{\mathrm{claim}}(X_i)$ and $\lambda_{\mathrm{crash}}(X_i)$ requires the additional restriction $\rho_{it}=1$ almost surely, which we do not impose. An actuarial precedent is provided by \citet{boucher2009accidents}, who model reported motor claims as a censored version of an underlying accident process when policyholders may refrain from reporting small accidents. Here $\rho_{it}$ is broader, allowing coverage, responsibility attribution and administrative recording to vary across crashes. A claim-frequency relativity need not match the corresponding crash-frequency relativity. External road-safety evidence reaches an insurance contrast through two mappings: from the study estimand to the target crash mechanism, and from crash occurrence to the recorded claim. The second mapping belongs to the observation process, not to the crash-mechanism DAG.

\begin{table}[H]
\centering
\small
\begin{tabularx}{\textwidth}{@{}p{2.0cm}p{2.5cm}X@{}}
\toprule
Symbol & Level & Role \\
\midrule
$X_i$ & policy / annual & observed rating information used by the insurer \\
$m_i$ & policy / annual & observed annual mileage \\
$M_i$ & policy / annual & number of trips or pre-specified trip segments in the policy period \\
$W_i$ & policy / slowly varying & latent experience, heterogeneity, activity, mobility, exposure and vehicle states \\
$\Omega_{it}$ & opportunity & displayed persistent and transient state relevant to opportunity $t$ \\
$U_i$ & annual latent history & $W_i$ plus the random-length sequence of opportunity states up to $M_i$ \\
$Y_{it}$ & opportunity & crash-occurrence indicator \\
$J_{it}$ & opportunity / observation & recorded liability-claim contribution generated after a crash \\
$\rho_{it}$ & crash / observation & conditional mean recorded-claim contribution given a crash \\
$Q_x$ & annual law & conditional law $\mathcal L(M_i,U_i\mid X_i=x)$ \\
\bottomrule
\end{tabularx}
\caption{Core notation and the scale at which each object is defined.}
\label{tab:notation}
\end{table}

\section{CAUSAL STRUCTURE FOR RISK CHARACTERIZATION}
\label{sec:escdag}

\subsection{Target outcome and graph scale}

The structural target is \emph{crash occurrence} during a driving opportunity, not culpability or severity conditional on a crash. Crash occurrence, responsibility attribution, injury severity and recorded insurance claims are distinct outcomes. Responsibility is deliberately handled downstream in the crash-to-claim observation process rather than folded into $Y_{it}$. Conditioning on severe crashes can create selection or collider bias when a determinant of the exposure also affects entry into the analysed sample \cite{dufournet2016selection}. Studies of culpability, injury or fatal crashes can still inform the graph, but their estimands are not substituted for crash occurrence without an explicit mapping.

The graph is a time-ordered snapshot of one driving opportunity. Fatigue, attention, speed and distraction can change within a trip and can affect one another at later instants; those lagged relations are compressed into an acyclic ordering at the chosen scale. Interventions with within-trip or cross-trip feedback require a dynamic DAG or structural time-series model.

\subsection{Evidence-informed graph construction}

We use the ESC-DAG logic of \citet{ferguson2020escdag} to organize heterogeneous road-safety evidence around candidate relations. The retained material contains 72 study--edge records. It is a targeted, iterative evidence map, not a systematic review or a random sample of the literature. Its purpose is to make the proposed graph traceable and revisable; absence from the map does not imply absence of a mechanism.

Candidate relations were reviewed in a fixed order: target estimand and temporal scale; temporal ordering and mechanistic plausibility; design-specific threats such as confounding, selection and measurement error; cross-design consistency; and graph status. Quantitative reusability was judged separately. Q1 estimates require relatively limited bridge assumptions, Q2 estimates require a more substantive mapping, Q3 evidence is structural only, and P records inform the predictive layer. One author conducted the extraction and adjudication, so no inter-rater statistic is available. We therefore treat $G_0$ as a provisional working graph, not a definitive causal map.

Most records inform structure rather than numerical bounds. The Appendix gives the extraction schema, study-level tables, adjudication rationales and graph-sensitivity alternatives. Only the Gomes-Franco mediation estimate enters the worked age geometry, where the cross-study bridge is stated explicitly and treated as a sensitivity assumption.

\section{A MULTI-RESOLUTION REPRESENTATION OF RISK GENERATION AND OBSERVATION}
\label{sec:integrated}
\label{sec:model}

\subsection{Persistent, contextual and transient components}

Let $R_i$ denote accumulated driving experience, $H_i$ residual persistent heterogeneity, $A_i$ a stable activity/use profile, $G_i$ the recurrent mobility environment, and $E_i$ latent exposure intensity. Annual mileage $m_i$ measures part of $E_i$ but is not equated with the number of opportunities $M_i$. Collect the slowly varying states as
\[
 W_i=(R_i,H_i,A_i,G_i,E_i,V_i^{\mathrm{perf}},V_i^{\mathrm{safe}}).
\]
The local state $\Omega_{it}$ combines the persistent coordinates that enter opportunity $t$ with the transient variables $(C_{it},S_{it},B_{it},K_{it})$. The remaining elements of $W_i$ govern opportunity count, context assignment, or upstream candidate paths.

The predictive layer is
\begin{equation}
  W_i\mid X_i\sim q_W(\cdot\mid X_i).
  \label{eq:measurementlayer}
\end{equation}
This is not a causal factorization. It describes what annual rating information says about slowly varying latent states. Age and licence tenure inform $R_i$; past claims inform $H_i$; declared use informs $A_i$; territory informs $G_i$; mileage informs $E_i$; and vehicle information bears on physical vehicle mechanisms. These are measurement or prediction links, shown as dashed arrows in Figure~\ref{fig:dag}.

Conditional on $W_i$, the solid arrows define the baseline graph $G_0$. At block level,
\begin{align}
M_i\mid(E_i,A_i,G_i) &\sim \kappa_M(\cdot\mid E_i,A_i,G_i), \\
C_{it} &\sim \kappa_C(c\mid A_i,G_i), \\
S_{it} &\sim \kappa_S(s\mid R_i,C_{it}), \\
B_{it} &\sim \kappa_B(b\mid R_i,H_i,C_{it},S_{it}), \\
K_{it} &\sim \kappa_K(k\mid C_{it},B_{it},V_i^{\mathrm{safe}}), \\
Y_{it} &\sim \mathrm{Bernoulli}\{\pi_Y(K_{it})\} \qquad \text{(baseline proximal-state restriction)}.
\label{eq:factorization}
\end{align}
The last line imposes \emph{proximal-state sufficiency}: conditional on $K_{it}$, the other displayed variables have no direct arrow to $Y_{it}$. This is a modeling restriction, not an empirical result. The annual identities do not depend on it, and the sensitivity family allows selected direct parents of $Y_{it}$. No independence across opportunities is imposed.

Using an infinite potential sequence only as notation, the annual latent history is the random-length object
\[
 U_i=\left(W_i,\{C_{it},S_{it},B_{it},K_{it}\}_{t=1}^{M_i}\right).
\]
The law $Q_x=\mathcal L(M_i,U_i\mid X_i=x)$ is induced by $q_W$, $\kappa_M$ and the within-opportunity kernels above. The block representation is coarser than the component-level graph: $C_{it}$ collects road, traffic, weather, passenger and timing context; $S_{it}$ transient state; and $B_{it}$ immediate behaviour. Two exclusions in $G_0$ are worth noting: $H_i\to S_{it}$ and $R_i\to C_{it}$. They are treated as sensitivity edges rather than scientific impossibilities.

\subsection{The DAG and the actuarial observation layer}

\begin{figure}[H]
\centering
\resizebox{0.99\textwidth}{!}{%
\begin{tikzpicture}[
  x=1cm,y=1cm,>=Latex,
  causal/.style={draw,rounded corners,align=center,minimum height=7mm,minimum width=18mm,fill=white,font=\scriptsize},
  latent/.style={draw,dashed,rounded corners,align=center,minimum height=7mm,minimum width=22mm,fill=gray!8,font=\scriptsize},
  rating/.style={draw,dotted,rounded corners,align=center,minimum height=7mm,minimum width=18mm,fill=gray!4,font=\scriptsize},
  info/.style={-{Latex},dashed,line width=0.65pt,gray!55},
  carrow/.style={-{Latex},line width=0.85pt},
  uarrow/.style={-{Latex},densely dotted,line width=1.05pt,black},
  layer/.style={draw=none,font=\scriptsize\itshape,gray}
]

\node[rating] (sex) at (-0.7,10.2) {Sex};
\node[rating] (age) at (1.1,10.2) {Age};
\node[rating] (tenure) at (3.0,10.2) {Licence\\tenure};
\node[rating] (mileage) at (5.1,10.2) {Annual\\mileage};
\node[rating] (use) at (7.2,10.2) {Declared\\use};
\node[rating] (territory) at (9.5,10.2) {Territory};
\node[rating] (history) at (11.9,10.2) {Claims /\\offences};
\node[rating] (vrating) at (14.7,10.2) {Vehicle\\rating info};

\node[latent] (experience) at (2.4,8.2) {Accumulated\\experience $R_i$};
\node[latent] (H) at (5.6,8.2) {Persistent\\state $H_i$};
\node[latent] (A) at (8.5,8.2) {Activity / use\\profile $A_i$};
\node[latent] (G) at (11.4,8.2) {Mobility\\environment $G_i$};
\node[causal] (vperf) at (14.3,8.2) {Vehicle\\performance};
\node[causal] (adas) at (16.9,8.2) {Vehicle safety\\ / ADAS};

\node[latent] (E) at (1.0,6.0) {Exposure\\intensity $E_i$};
\node[causal] (pass) at (3.5,6.0) {Passengers};
\node[causal] (night) at (5.7,6.0) {Night};
\node[causal] (road) at (8.3,6.0) {Road type};
\node[causal] (traffic) at (10.8,6.0) {Traffic};
\node[causal] (weather) at (13.2,6.0) {Weather};

\node[causal] (distract) at (3.5,3.8) {Distraction};
\node[causal] (fatigue) at (5.7,3.8) {Fatigue};
\node[causal] (attention) at (8.3,3.8) {Attention /\\reaction};
\node[causal] (viol) at (10.8,3.8) {Violations};
\node[causal] (speed) at (13.2,3.8) {Speed};

\node[causal] (errors) at (8.3,1.6) {Errors};
\node[causal] (conflict) at (13.8,1.6) {Critical\\conflict};
\node[causal] (crash) at (16.8,1.6) {Crash\\occurrence};

\draw[carrow] (experience) to[bend right=12] (attention);
\draw[carrow] (experience) to[bend right=22] (errors);
\draw[carrow] (A) -- (pass);
\draw[carrow] (A) -- (night);
\draw[carrow] (A) to[bend left=10] (road);
\draw[carrow] (G) -- (road);
\draw[carrow] (G) -- (traffic);
\draw[carrow] (G) -- (weather);

\draw[uarrow] (pass) -- (distract);
\draw[carrow] (night) -- (fatigue);
\draw[carrow] (distract) -- (attention);
\draw[carrow] (fatigue) -- (attention);
\draw[carrow] (attention) -- (errors);
\draw[carrow] (H) to[bend right=14] (viol);
\draw[carrow] (H) to[bend left=13] (speed);
\draw[carrow] (H) to[bend right=24] (errors);
\draw[carrow] (road) to[bend right=8] (speed);
\draw[carrow] (traffic) -- (speed);
\draw[uarrow] (vperf) -- (speed);

\draw[uarrow] (errors) to[bend right=8] (conflict);
\draw[uarrow] (viol) -- (conflict);
\draw[carrow] (speed) -- (conflict);
\draw[carrow] (weather) to[bend left=8] (conflict);
\draw[carrow] (adas) to[bend left=10] (conflict);
\draw[carrow] (conflict) -- (crash);

\draw[info] (age) -- (experience);
\draw[info] (tenure) -- (experience);
\draw[info] (sex) -- (H);
\draw[info] (sex) to[bend right=12] (A);
\draw[info] (mileage) -- (E);
\draw[info] (use) -- (A);
\draw[info] (territory) -- (G);
\draw[info] (history) -- (H);
\draw[info] (vrating) -- (vperf);
\draw[info] (vrating) -- (adas);

\node[layer,anchor=west] at (18.3,10.2) {actuarial observations};
\node[layer,anchor=west] at (18.3,8.2) {persistent / assignment};
\node[layer,anchor=west] at (18.3,6.0) {trip context};
\node[layer,anchor=west] at (18.3,3.8) {state / behaviour};
\node[layer,anchor=west] at (18.3,1.0) {proximal outcome};

\end{tikzpicture}%
}
\caption{The integrated DAG with explicit observation and assignment layers. Solid black arrows represent retained structural relations, heavier dotted arrows represent relations adjudicated as uncertain, and lighter dashed arrows represent prediction or measurement links from actuarial variables. Age and licence tenure inform accumulated experience $R_i$; declared use measures an activity/use profile $A_i$; territory measures a mobility-environment profile $G_i$; mileage measures exposure intensity $E_i$; and claims/offence history measures persistent heterogeneity $H_i$. The per-opportunity crash graph is a time-ordered snapshot aggregated over the number and composition of driving opportunities; dashed links are not part of the causal DAG.}
\label{fig:dag}
\end{figure}
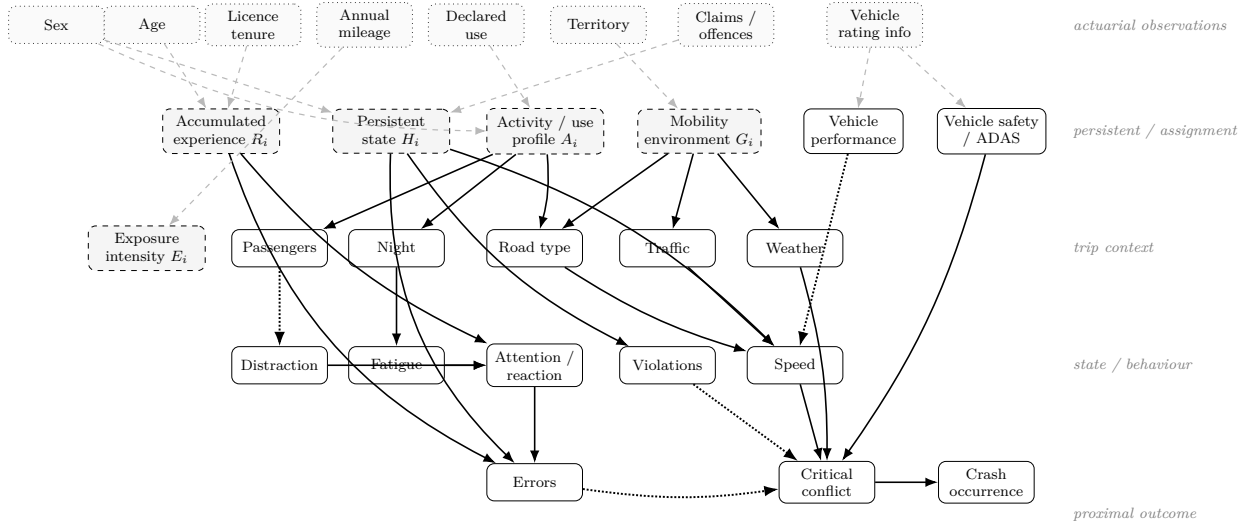

Figure~\ref{fig:dag} separates three kinds of relation. Solid arrows define the structural core, densely dotted arrows are plausible but uncertain, and dashed arrows belong only to the observation layer. Absence of a solid arrow is a maintained structural restriction. D-separation claims are therefore conditional on the variables shown and on the assumed absence of omitted common causes \cite{greenland1999causal,pearl2009causality}. Dashed links are not used to derive adjustment sets or $do$-operator effects. Declared use measures $A_i$, territory measures $G_i$, and claims history measures $H_i$; none is inserted into the crash mechanism simply because it predicts claims.

\subsection{Graph uncertainty and nested structural models}
\label{sec:graphsensitivity}

Let $G_0$ contain the solid arrows in Figure~\ref{fig:dag}, $G_+$ add the dotted candidate relations, and $\mathbb G$ include the additional sensitivity graphs in the Appendix. Dashed actuarial bridges belong to none of these causal graphs.

\paragraph{Proposition 1 (nested-graph monotonicity).}
Let $G_0$ and $G_1$ be nonparametric Markov models with the same nodes, state spaces, temporal ordering, observation class $\mathcal R_x$, and claim functional. If $G_1$ contains every arrow in $G_0$ and differs only by removing conditional-independence restrictions, then
\begin{equation}
 \mathcal Q_{G_0}(x)\subseteq\mathcal Q_{G_1}(x)
 \quad\Longrightarrow\quad
 \mathcal F_{G_0}(r)\subseteq\mathcal F_{G_1}(r).
 \label{eq:graphmonotonicity}
\end{equation}
\emph{Proof.} Every law and observation mapping admissible under $G_0$ remains admissible under $G_1$, while the claim-ratio restriction is unchanged. $\square$ The result applies only to this nested setting; it says nothing about changes in node definition, parameterization, observation model, or transport restrictions.

For a finite graph family,
\begin{equation}
 \mathcal F_{\mathbb G}(r)
 =\bigcup_{G\in\mathbb G}\mathcal F_G(r),
 \qquad
 \mathcal F_{\mathbb G,\mathrm{ext}}(r)
 =\bigcup_{G\in\mathbb G}
 \bigl(\mathcal F_G(r)\cap\mathcal F_{\mathrm{RS},G}\bigr),
 \label{eq:graphunion}
\end{equation}
where $\mathcal F_{\mathrm{RS},G}$ contains only external restrictions that are meaningful under graph $G$. The empirical illustration does not compute these full-DAG sets. Its later three-block geometry is a separate display model, not a projection of $\mathcal F_G(r)$.

\subsection{From trip-level causation to annual risk}

The annual mean admits an exact accounting decomposition without assuming independent opportunities; this algebra should not be read as a causal separation of exposure from composition. To keep the zero-opportunity case explicit, define for $M_i>0$
\[
 \bar p_i = \frac{1}{M_i}\sum_{t=1}^{M_i}p_{it},
 \qquad
 \pi_+(X_i)=\Pr(M_i>0\mid X_i).
\]
Because the random sum in Eq.~\eqref{eq:central} is zero when $M_i=0$,
\begin{equation}
\begin{aligned}
 \lambda_{\mathrm{crash}}(X_i)
 =\pi_+(X_i)\Bigl[
 &\mathbb E(M_i\mid X_i,M_i>0)\,
  \mathbb E(\bar p_i\mid X_i,M_i>0)\\
 &+\operatorname{Cov}(M_i,\bar p_i\mid X_i,M_i>0)
 \Bigr].
\end{aligned}
\label{eq:quantityquality}
\end{equation}
If $\Pr(M_i>0\mid X_i)=1$, the leading factor is one and this reduces to the familiar covariance decomposition. The covariance term is an exact accounting term, not an estimated mechanism: it records dependence between how much a driver is exposed and the risk composition of that exposure.

Observed mileage need not behave as a linear offset in an annual frequency model. Drivers with different mileage can differ in route, time, traffic and behavioural composition, so exposure quantity may be associated with average opportunity risk. Elvik \cite{elvik2023mileage} documents a substantially sublinear aggregate relation between annual distance and accident involvement. That relation constrains quantity and composition jointly without selecting a mechanism.

\section{RATING INFORMATION AND RISK OBSERVATION}
\label{sec:rating}

Rating variables are treated mainly as information about latent mechanisms. Age is a useful example. For two age groups $a$ and $a_0$, the structural crash relativity is
\begin{equation}
 RR^{\mathrm{crash}}_{\mathrm{age}}(a;a_0)
 =
 \frac{\mathbb{E}[\sum_{t=1}^{M_i} p_{it}\mid \mathrm{Age}_i=a]}
 {\mathbb{E}[\sum_{t=1}^{M_i} p_{it}\mid \mathrm{Age}_i=a_0]}.
 \label{eq:rrage}
\end{equation}
The insurance analogue replaces the crash sum by $\sum_t p_{it}\rho_{it}$. These ratios coincide only under restrictions on the claim-observation process. Licence tenure is closer to accumulated experience $R_i$ than age alone, but age and tenure still do not separate maturation, cohort and time-since-licensure \cite{mccartt2003experience,gulliver2013learner,curry2015experience,ehsani2020learner}. The \texttt{freMTPL2freq} illustration contains age but not licence tenure.

Sex and claims history also belong to the observation layer in the core model. Sex can carry information about mileage, time of day and speed distributions \cite{massie1997gender,regev2018agegender,ayuso2016gender,guillen2021speeding}; a single sex-to-crash arrow would hide these routes. Claims history has a clear temporal ordering:
\begin{equation}
\begin{aligned}
 H_i,R_i,\ldots &\longrightarrow Y_{i,<t}
 \longrightarrow J_{i,<t}
 \longrightarrow \mathrm{PastClaims}_i,\\
 H_i &\longrightarrow B_{it}\longrightarrow K_{it}\longrightarrow Y_{it}.
\end{aligned}
 \label{eq:claimhistory}
\end{equation}
The dashed $\mathrm{PastClaims}_i\to H_i$ arrow therefore denotes statistical updating, not physical causation. Bonus--malus inherits the same historical character.

Declared use and territory are linked to stable bridge objects rather than directly to crash: use informs the activity profile $A_i$, while territory informs the mobility environment $G_i$ \cite{elias2010activity,robb2008work,newnam2022systems,chipman1993exposure,blatt1998residence,lee2014residence,ma2018context,guillen2024weekly,shi2017territorial}. Vehicle information is different because collision-avoidance technology can act directly through the conflict branch. Broad vehicle categories nevertheless mix physical effects with severity and driver--vehicle selection \cite{hoye2019vehicle,mccartt2017power,keall2013performance}. The graph therefore separates safety and performance mechanisms from the coarse vehicle label.

\section{COMPATIBILITY AND STRUCTURAL UNCERTAINTY}
\label{sec:partial}

\subsection{An uncertainty inventory for mechanistic interpretation}

The set-valued approach follows the partial-identification tradition \citep{manski2003partial,tamer2010partial}. Here it is used to keep four sources of uncertainty distinct. \emph{Statistical uncertainty} concerns estimation of an annual contrast. \emph{Structural uncertainty} concerns admissible graphs and latent mechanism laws. \emph{Observation uncertainty} concerns the crash-to-claim map $\rho$. \emph{Transport uncertainty} concerns the correspondence between an external study and the target population, outcome and estimand. A narrow confidence interval for a claim relativity addresses only the first.

The set $\mathcal F_G(r)$ collects structural and observation laws compatible with an annual contrast under graph $G$. By contrast, the $\delta$-indexed envelopes below are sensitivity sets conditional on analyst-specified discrepancy bounds; $\delta$ is not learned from the French portfolio. The numerical examples are lower-dimensional than the DAG and do not estimate causal effects, recover the full mechanism distribution, or validate the graph. The age simplex is a separate bookkeeping construction.

This use of sensitivity differs from omitted-variable robustness or E-values, which quantify unmeasured confounding for a fixed estimand \citep{cinelli2020sensitivity,vanderweele2017evalue}. Here $\delta$ stresses a proposed cross-study correspondence, while graph sensitivity is handled by changing the structural model. Keeping the two separate avoids hiding model-form and transport assumptions inside one number.

\subsection{External evidence as constraints on structural uncertainty}

External evidence yields a numerical constraint only after its estimand has been related to a target quantity. Formal transportability can sometimes derive that relation from explicit invariance assumptions \citep{bareinboim2013transport,pearlbareinboim2014external}; we do not claim such a formula for the heterogeneous studies used here. We instead separate a \emph{bridge specification} from residual discrepancy. If the bridge is not defensible, the study remains qualitative evidence.

Reported effects remain on their native scale. For a ratio estimate $\widehat\eta_{ej}$ with confidence limits $(L_{ej},U_{ej})$, define
\[
 z_{ej}=\log(\widehat\eta_{ej}),
 \qquad
 s_{ej}\simeq\frac{\log(U_{ej})-\log(L_{ej})}{2\times1.96}.
\]
Other estimands require design-specific transformations; the resulting study summaries are not assumed exchangeable.

Let $\zeta_{ej}$ denote the bridge assumptions required to relate study $j$ to target relation $e$: for example, outcome-scale conversion, selection into the analysed population, effect-scale correspondence, or population invariance. Write $T_{d_j,o_j,\mathsf{pop}_j}(\theta_e;\zeta_{ej})$ for the study estimand implied by target quantity $\theta_e$ under those assumptions. Then
\begin{equation}
 z_{ej}=T_{d_j,o_j,\mathsf{pop}_j}(\theta_e;\zeta_{ej})
       +\Delta_{ej}+\varepsilon_{ej},
 \qquad
 \varepsilon_{ej}\sim N(0,s_{ej}^2),
 \label{eq:transport}
\end{equation}
where $\Delta_{ej}$ is residual mismatch after the bridge is specified. A discrepancy bound cannot make fundamentally different estimands commensurate. Mediation parameters require their own identifying assumptions \citep{imai2010mediation}, and an odds ratio conditional on crash involvement is not a rate ratio for annual claims.

For $\boldsymbol\delta_e=(\delta_{e1},\ldots,\delta_{eJ_e})$ with $|\Delta_{ej}|\le\delta_{ej}$ and bridge specifications $\boldsymbol\zeta_e=(\zeta_{e1},\ldots,\zeta_{eJ_e})$, define the sensitivity-feasible set
\begin{equation}
 \Pi_e(\boldsymbol\delta_e;\boldsymbol\zeta_e)=\left\{\theta_e:
 |z_{ej}-T_{d_j,o_j,\mathsf{pop}_j}(\theta_e;\zeta_{ej})|
 \le 1.96s_{ej}+\delta_{ej}
 \quad\text{for every }j
 \right\}.
 \label{eq:edgeenvelope}
\end{equation}
This is a sensitivity-feasible set, not a joint 95\% confidence region. With heterogeneous studies it may be disconnected or empty; emptiness means that the evidence and imposed bridge restrictions are mutually incompatible.

For a set of target relations $\mathcal E^\star$, let $\vartheta=(\theta_e:e\in\mathcal E^\star)$ and define
\begin{equation}
\begin{aligned}
 \Pi_{\mathcal E^\star}(\boldsymbol\delta;\boldsymbol\zeta)
 =\bigl\{\vartheta:\;&
 \theta_e\in\Pi_e(\boldsymbol\delta_e;\boldsymbol\zeta_e)
 \quad\text{for every }e\in\mathcal E^\star,\\
 &\vartheta\text{ satisfies the stated cross-relation structural restrictions}
 \bigr\}.
\end{aligned}
 \label{eq:jointenvelope}
\end{equation}
No independence assumption is implied. If a displayed bookkeeping term is $c_k=b_k(\vartheta)$, its sensitivity bounds are
\begin{equation}
 \ell_k(\boldsymbol\delta;\boldsymbol\zeta)=
 \inf_{\vartheta\in\Pi_{\mathcal E^\star}(\boldsymbol\delta;\boldsymbol\zeta)} b_k(\vartheta),
 \qquad
 u_k(\boldsymbol\delta;\boldsymbol\zeta)=
 \sup_{\vartheta\in\Pi_{\mathcal E^\star}(\boldsymbol\delta;\boldsymbol\zeta)} b_k(\vartheta).
 \label{eq:blockprojection}
\end{equation}
For a fixed bridge, larger discrepancy bounds weakly enlarge the feasible set by construction. The worked example uses one study and an intentionally strong bridge; it is a stress test, not an application of formal transportability.

\subsection{Predictive contrasts as constraints, not mechanism estimates}

Let $r_{\mathrm{claim}}(x)$ be a claim-frequency relativity for $X=x$ relative to $x_0$, and recall the annual law
\[
 Q_x=\mathcal L(M_i,U_i\mid X_i=x),
\]
with the slowly varying marginal $q_W(\cdot\mid x)$ specified in Eq.~\eqref{eq:measurementlayer}. For a given annual history $(M,U)$, define the conditional claim sum
\begin{equation}
 \mu_{\mathrm{claim}}(M,U;x)
 =\sum_{t=1}^{M} p_t(U)\,\rho_t(U,x),
 \label{eq:annualclaimfunctional}
\end{equation}
where $p_t(U)$ is the crash probability and $\rho_t$ the conditional mean claim contribution. The observed relativity is
\begin{equation}
 r_{\mathrm{claim}}(x)
 =
 \frac{\mathbb{E}_{Q_x}\{\mu_{\mathrm{claim}}(M,U;x)\}}
      {\mathbb{E}_{Q_{x_0}}\{\mu_{\mathrm{claim}}(M,U;x_0)\}}.
 \label{eq:inverse}
\end{equation}
Setting $\rho_t\equiv1$ would recover a crash-frequency functional, a restriction not justified by claim data.

For graph $G$, let $\mathcal Q_G(x)$ be the set of annual laws $Q_x$ induced by structural models compatible with $G$, its temporal ordering and the maintained no-unmodelled-common-cause assumptions. Let $\rho_x(t,U)$ be a measurable crash-to-claim conditional-mean mapping and let $\mathcal R_x$ denote its admissible class. Define the claim functional
\begin{equation}
 \mu(Q_x,\rho_x)
 =
 \mathbb E_{Q_x}\!\left[
   \sum_{t=1}^{M}p_t(U)\rho_x(t,U)
 \right].
 \label{eq:claimfunctionalQ}
\end{equation}
The graph-specific compatibility set is then
\begin{equation}
\begin{aligned}
 \mathcal F_G(r)=\biggl\{(Q_x,Q_{x_0},\rho_x,\rho_{x_0}):\;&
 Q_x\in\mathcal Q_G(x),\quad Q_{x_0}\in\mathcal Q_G(x_0),\\
 &\rho_x\in\mathcal R_x,\quad \rho_{x_0}\in\mathcal R_{x_0},\\
 &\frac{\mu(Q_x,\rho_x)}
 {\mu(Q_{x_0},\rho_{x_0})}=r
 \biggr\}.
\end{aligned}
 \label{eq:Fset}
\end{equation}
Road-safety evidence may restrict crash mechanisms and study-to-target mappings under graph $G$, but it does not identify $\rho$. Denote the graph-specific external restrictions by $\mathcal F_{\mathrm{RS},G}$. Then
\begin{equation}
 \mathcal F_{G,\mathrm{ext}}(r)
 =\mathcal F_G(r)\cap\mathcal F_{\mathrm{RS},G}.
 \label{eq:idset}
\end{equation}
Equation~\eqref{eq:graphunion} extends this definition across $\mathbb G$. Precise prediction of a tariff contrast can therefore coexist with weak restrictions on both crash pathways and the crash-to-claim bridge.

\paragraph{Proposition 2 (no crash-ratio bound under unrestricted observation).}
Suppose the observed group-specific claim means $\lambda_{\mathrm{claim}}(x)$ and $\lambda_{\mathrm{claim}}(x_0)$ are positive and the admissible observation class allows arbitrary positive constant mappings $\rho_x$ and $\rho_{x_0}$. Then, for any positive candidate crash means $\lambda_{\mathrm{crash}}(x)$ and $\lambda_{\mathrm{crash}}(x_0)$, choosing
\begin{equation}
 \rho_x=\frac{\lambda_{\mathrm{claim}}(x)}{\lambda_{\mathrm{crash}}(x)},
 \qquad
 \rho_{x_0}=\frac{\lambda_{\mathrm{claim}}(x_0)}{\lambda_{\mathrm{crash}}(x_0)}
 \label{eq:rho_nonid}
\end{equation}
reproduces the observed claim means. Hence a claim-frequency ratio alone gives no non-trivial bound on the crash-frequency ratio when the observation map is unrestricted. Informative bounds require restrictions on $\mathcal R_x$, such as bounded recording rates, cross-group invariance, or linked crash--responsibility--claim data.

\paragraph{Operational boundary.}
Computing the full set would additionally require a finite graph family, a bounded or estimable observation class $\mathcal R_x$, bridge specifications for external evidence, and a computational class for $Q_x$. With those choices, $\mathcal F_G(r)$ becomes a feasibility problem for optimization, simulation, or Bayesian computation. We do not propose a general solver here.

For a low-dimensional illustration, impose the separate accounting relation $\mathbf 1^\top c=L=\log r$ on $K$ displayed log terms $c=(c_1,\ldots,c_K)$. Neither the DAG nor Eq.~\eqref{eq:inverse} implies this additivity, and the terms are not a unique path decomposition. The unrestricted display set is
\begin{equation}
 \mathcal{C}(L)=\{c\in\mathbb{R}^K:\mathbf{1}^{\top}c=L\}.
 \label{eq:hyperplane}
\end{equation}
which is unbounded for $K\ge2$. Imposing $c_k\ge0$ gives the simplex
\begin{equation}
 \mathcal{C}_{+}(L)=\mathcal{C}(L)\cap\mathbb{R}_{+}^{K},
 \label{eq:simplexgeneral}
\end{equation}
while analyst-imposed or externally motivated display bounds $\ell_k(\boldsymbol\delta;\boldsymbol\zeta)\le c_k\le u_k(\boldsymbol\delta;\boldsymbol\zeta)$ produce the convex polytope
\begin{equation}
 \mathcal{C}_{\mathrm{ext}}(L;\boldsymbol\delta)=
 \mathcal{C}(L)\cap\{c:\ell_k(\boldsymbol\delta;\boldsymbol\zeta)\le c_k\le u_k(\boldsymbol\delta;\boldsymbol\zeta),\ k=1,\ldots,K\}.
 \label{eq:polytope}
\end{equation}
This simplex is a conditional display model. Non-negativity is a substantive sign restriction; allowing signed terms returns Eq.~\eqref{eq:hyperplane}. The later figure asks only which allocations survive an illustrative bound within this coarse representation.

\subsection{Two quantitative submodels}
\label{sec:quantitative}

Two limited examples illustrate the identification problem. Mileage gives an aggregate accounting identity; age shows how a claim contrast changes with conditioning and then uses the toy compatibility display.

\subsubsection{Mileage: an aggregate rate constraint per unit distance}

Let $m$ denote annual mileage. The object here is the \emph{marginal aggregate} accident relation in the mileage literature, not the conditional insurance mean $\lambda_{\mathrm{claim}}(m,Z)$. Write $\lambda_{\mathrm{agg}}(m)$ and define
\begin{equation}
  \bar p_{\mathrm{agg}}(m)=\frac{\lambda_{\mathrm{agg}}(m)}{m},
  \qquad\text{so that}\qquad
  \lambda_{\mathrm{agg}}(m)=m\,\bar p_{\mathrm{agg}}(m).
  \label{eq:mileagefactor}
\end{equation}
Taking log derivatives of this identity gives
\begin{equation}
  \eta_{\lambda,m}=1+\eta_{\bar p,m},
  \qquad
  \eta_{\lambda,m}=\frac{d\log\lambda_{\mathrm{agg}}(m)}{d\log m}.
  \label{eq:mileageelasticity}
\end{equation}
More generally, an aggregate power approximation $\lambda_{\mathrm{agg}}(m)\propto m^{\alpha}$ implies
\begin{equation}
  \eta_{\bar p,m}=\alpha-1.
  \label{eq:compositionelasticity}
\end{equation}
Elvik's synthesis uses $\alpha\approx1/2$ as a working approximation for the studies considered \cite{elvik2023mileage}, not as a universal exponent. A sensitivity check with $\alpha\in\{0.4,0.5,0.6\}$ gives fourfold accident multipliers of 1.74, 2.00 and 2.30 and aggregate-rate-per-mile multipliers of 0.44, 0.50 and 0.57.

The declining aggregate rate per mile is an arithmetic consequence of sublinearity, not evidence that additional mileage protects a driver. Route and time composition, selection, experience and unobserved heterogeneity may all contribute \cite{elvik2023mileage,janke1991mileage,langford2006lowmileage,antin2017lowmileage}.

If the aggregate accident rate per mile is factorized multiplicatively into mechanism blocks for bookkeeping, the corresponding elasticities add:
\begin{equation}
 \eta_{\bar p,m}=\eta_{\mathrm{road}}+\eta_{\mathrm{time}}+\eta_{\mathrm{traffic}}+\eta_{\mathrm{experience/selection}}+\eta_{\mathrm{other}}.
 \label{eq:mileageblocks}
\end{equation}
Only the sum in Eq.~\eqref{eq:mileageblocks} is fixed by this bookkeeping relation; the DAG does not identify its components.

\begin{figure}[H]
\centering
\includegraphics[width=0.78\textwidth]{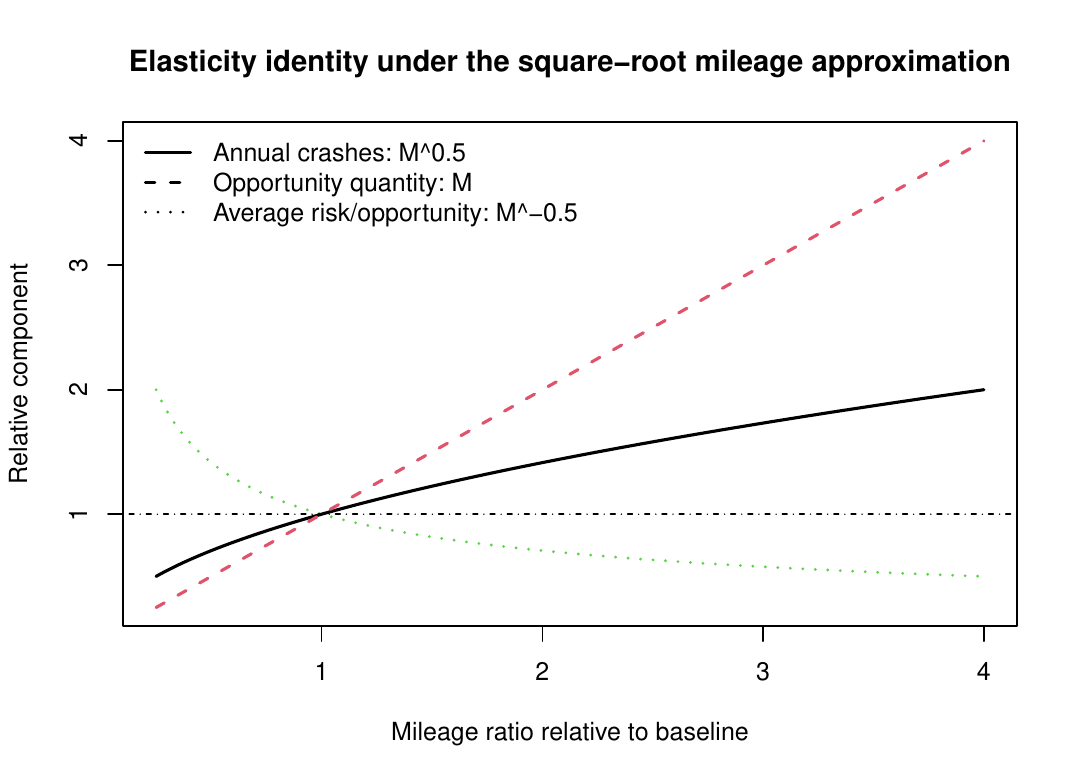}
\caption{Aggregate elasticity identity under the working approximation $\lambda_{\mathrm{agg}}(m)\propto m^{1/2}$. Observed mileage grows linearly by definition of the horizontal exposure scale, while the associated accident count grows as $m^{1/2}$ and the ratio $\lambda_{\mathrm{agg}}(m)/m$ declines as $m^{-1/2}$. The declining ratio is an aggregate constraint, not an identified causal composition effect.}
\label{fig:mileageelasticity}
\end{figure}

\subsubsection{Age: the rating relativity depends on the information set}

The age illustration uses the open French motor third-party-liability portfolio \texttt{freMTPL2freq} from \texttt{CASdatasets} (677,991 records) \cite{dutang2026casdatasets,noll2020fremtpl}. Restricting to $0<\mathrm{Exposure}\leq1$ leaves 676,767 records. We fit claim counts by Poisson pseudo-maximum likelihood with log exposure as offset. Age is grouped into nine bands with 40--49 as reference; the vehicle/geographic specification also includes vehicle age, power, fuel, area, region and density deciles. The computational appendix records the exact coding.

Counts and exposures are aggregated over identical covariate patterns before fitting. The HC0 Huber--White covariance is computed on these aggregated observations; it is not a cluster sandwich over contract-level residuals. Because covariates are identical within each cell, aggregation preserves the Poisson score and the corresponding coefficient estimates. The computational appendix gives the construction. Here ``predictive'' refers to the conditional claim-risk functional, not to an out-of-sample forecasting comparison.

The vehicle/geographic model gives
\begin{equation}
 \widehat{RR}_{18--20:40--49}=3.388\quad (95\%\ \mathrm{CI}:\ 3.114,\ 3.686),
 \label{eq:agerelativity}
\end{equation}
compared with 3.182 (2.923--3.464) in the raw age model. Adding the medium bonus--malus discretisation gives 1.235 (1.129--1.352). These are different conditional claim contrasts. Bonus--malus is an experience-rating variable that compresses past claim history \citep{boucher2009accidents,boucher2022bonusmalus}; its predictive value does not make it a baseline causal adjustment variable. Table~\ref{tab:bmsensitivity} changes only the resolution of the recorded score: 25-, 10- and 5-point bins yield 8, 19 and 34 realised categories and relativities 1.381, 1.235 and 1.209.

\begin{table}[H]
\centering
\begingroup
\scriptsize
\setlength{\tabcolsep}{3pt}
\begin{tabular}{lrrrr}
\toprule
Bonus--malus specification & BM levels & Rating cells & $\widehat{RR}_{18--20:40--49}$ & 95\% CI \\
\midrule
Coarse & 8 & 94,840 & 1.381 & [1.264, 1.510] \\ 
Medium & 19 & 125,251 & 1.235 & [1.129, 1.352] \\ 
Fine & 34 & 153,155 & 1.209 & [1.104, 1.324] \\ 
\bottomrule
\end{tabular}

\endgroup
\caption{Sensitivity of the incremental 18--20 versus 40--49 age relativity to alternative fixed-width categorical resolutions of bonus--malus. The coarse, medium and fine specifications use bin widths 25, 10 and 5 respectively, anchored at 50. All specifications retain the same vehicle and geographic controls; only the resolution of bonus--malus changes.}
\label{tab:bmsensitivity}
\end{table}

\begin{figure}[H]
\centering
\resizebox{0.96\textwidth}{!}{%
\begin{tikzpicture}[>=Latex, node distance=12mm and 14mm,
  n/.style={draw,rounded corners,align=center,font=\scriptsize,inner sep=3pt},
  cond/.style={draw,double,rounded corners,align=center,font=\scriptsize,inner sep=3pt}]
\node[n] (age) {Age /\ tenure};
\node[n, right=of age] (latentbm) {$R_i,H_i$};
\node[n, right=of latentbm] (pastcrash) {past\ crash};
\node[n, right=of pastcrash] (pastclaim) {past liability\ claim};
\node[cond, right=of pastclaim] (bm) {bonus--malus\ conditioned on};
\node[n, below=of bm] (duration) {insurance-history\ duration};
\node[n, below=of pastcrash] (futurecrash) {future\ crash};
\node[n, right=of futurecrash] (futureclaim) {future liability\ claim};
\draw[->] (age) -- (latentbm);
\draw[->] (latentbm) -- (pastcrash);
\draw[->] (pastcrash) -- (pastclaim);
\draw[->] (pastclaim) -- (bm);
\draw[->] (duration) -- (bm);
\draw[->] (latentbm) -- (futurecrash);
\draw[->] (futurecrash) -- (futureclaim);
\end{tikzpicture}%
}
\caption{Schematic history sub-DAG for interpreting bonus--malus conditioning. The double border marks the variable conditioned on in the predictive model. The diagram does not assert that bonus--malus causes future crashes; it shows why conditioning on a downstream summary of past claims changes the target and can, under fuller graphs, induce associations among causes of the conditioned variable.}
\label{fig:bmdag}
\end{figure}
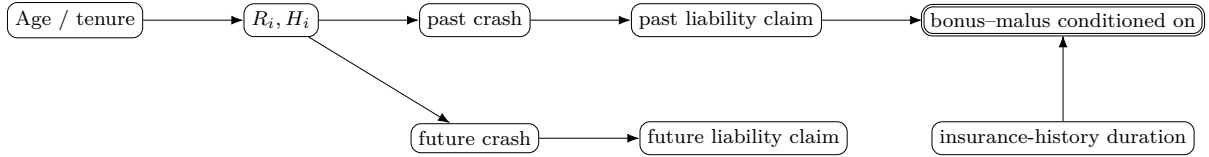

More generally, let $\mathcal Z$ denote the \emph{set of rating variables included in the fitted claim model}, rather than a random covariate value. For a log-link model without age interactions, write the fitted age relativity as
\begin{equation}
 r^{\mathrm{claim}}_{\mathrm{age}}(a;\mathcal Z)
 =\exp\{\beta_a^{(\mathcal Z)}-\beta_{a_0}^{(\mathcal Z)}\}.
 \label{eq:conditionalrating}
\end{equation}
Changing $\mathcal Z$ changes the predictive contrast before any mechanistic interpretation. Bonus--malus requires particular caution because it is not a baseline confounder. In the French system the coefficient evolves mechanically with claim-free insurance periods and responsible claims \cite{france2025bonusmalus}. Figure~\ref{fig:bmdag} summarizes the relevant history structure. Persistent heterogeneity and experience contribute to earlier crashes, earlier crash-to-claim realizations feed the recorded bonus--malus score, and insurance-history duration is another cause of that score. The same persistent states also contribute to future crash risk.

The value 1.235 should be read as a \emph{conditional predictive contrast}: under the fitted no-age-interaction log-link model, it compares fitted claim rates for the two age groups at the same included vehicle, geographic and bonus--malus values. Conditioning changes which insurance histories are being compared. Because bonus--malus has several causes, it may also induce non-causal associations among them under fuller graphs \cite{greenland1999causal,pearl2009causality}. The score can proxy persistent heterogeneity and insurance duration, but it is not accumulated driving experience $R_i$, and these channels are not separated in the portfolio.

\begin{figure}[H]
\centering
\includegraphics[width=0.82\textwidth]{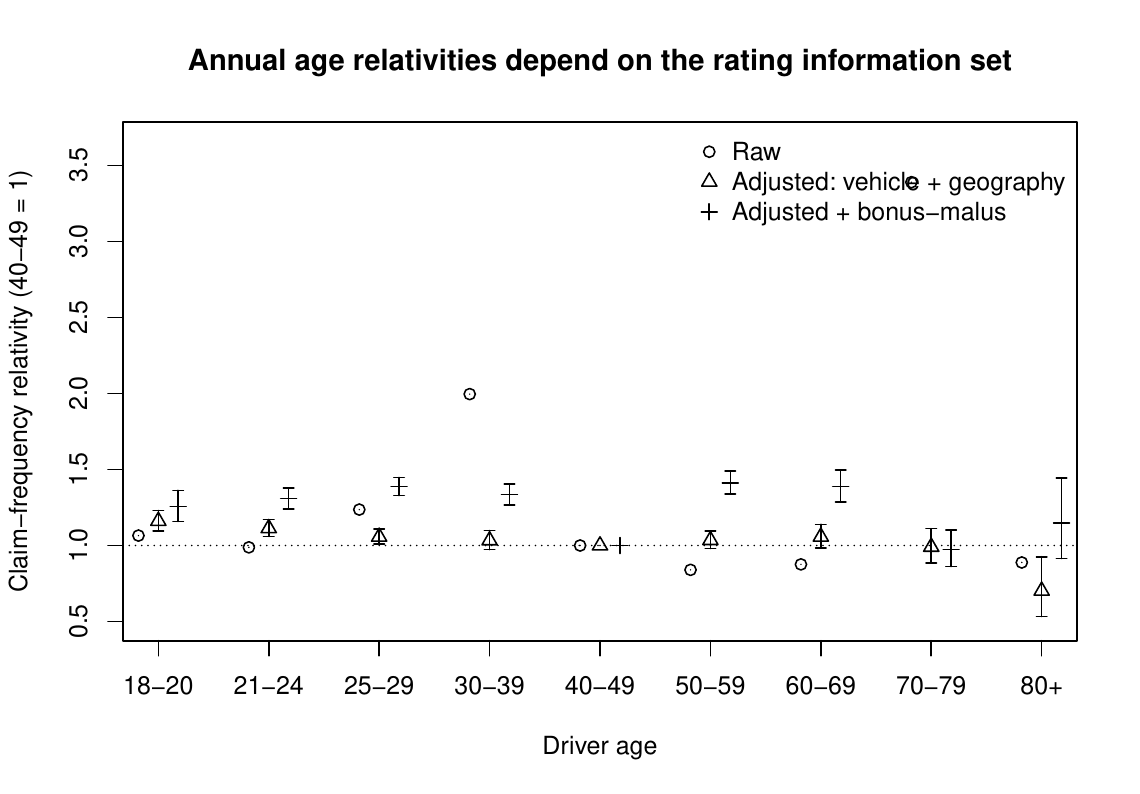}
\caption{Annual claim-frequency relativities by age in \texttt{freMTPL2freq}, with ages 40--49 as reference. Raw relativities are compared with the vehicle/geographic model and with the same model augmented by the medium bonus--malus discretisation (10-point bins; 19 realised levels). Intervals use the HC0 sandwich computed on aggregated rating-cell observations. Bonus--malus is an endogenous history variable, so the comparison describes alternative predictive conditioning sets; it is not a causal adjustment or an intervention on age.}
\label{fig:agerelativity}
\end{figure}

\subsubsection{Coarse compatibility sets for the age explanation}

The two conditioning sets imply different claim-scale constraints. Let
\[
 L_{\mathrm{base}}=\log(3.387706)=1.220153,
 \qquad
 L_{\mathrm{BM}}=\log(1.235315)=0.211326.
\]
The first uses vehicle/geographic information; the second also conditions on the medium bonus--malus specification. For the toy geometry, define $c_E$ (experience/human), $c_C$ (environment/vehicle/context), and a balance term $c_B$ that absorbs interactions, omitted mechanisms and crash-to-claim differences. The accounting identity is $c_E+c_C+c_B=L_z$. Without sign restrictions it defines an unbounded affine plane. The displayed triangle adds the substantive restriction $c_E,c_C,c_B\ge0$:
\begin{equation}
 \mathcal{C}_{+}(L_z)=
 \{(c_E,c_C,c_B)\in\mathbb{R}_+^3:c_E+c_C+c_B=L_z\},
 \qquad z\in\{\mathrm{base},\mathrm{BM}\}.
 \label{eq:agesimplex}
\end{equation}
Each point reproduces the corresponding claim relativity within this accounting model. Adding bonus--malus changes the target from $L_{\mathrm{base}}=1.220$ to $L_{\mathrm{BM}}=0.211$; it does not produce a causally adjusted age effect. The external-evidence exercise below therefore uses $L_{\mathrm{base}}$. Published evidence supports several age-related pathways \cite{curry2015experience,simonsmorton2005passengers,goodwin2012passengers,regev2018agegender,guillen2021speeding}, but the claim contrast does not determine their shares.

\subsubsection{A worked intersection with external road safety evidence}
\label{sec:workedintersection}

With no quantitative transport, the benchmark is simply $\mathcal C_+(L_{\mathrm{base}})$. To show how an external restriction would enter, we then impose a strong illustrative bridge. \citet{gomesfranco2020age} compare Spanish crash-involved drivers aged 18--24 with ages 35--44 and report a culpability OR of 2.15, with an indirect OR of 1.09 (95\% CI 1.08--1.10) through environmental and vehicle circumstances. This mediation parameter is tied to its own treatment, mediator, outcome and identifying assumptions \citep{imai2010mediation}; it is not a contribution to annual French claim frequency.

The source and target differ in effect scale, outcome, selection, age bands, population and claim observation. Formal transportability would require assumptions sufficient to derive the target estimand from source and target data \citep{bareinboim2013transport,pearlbareinboim2014external}; we do not have such a formula.

To make the sensitivity exercise transparent, let $\zeta_C^{\mathrm{id}}$ denote the \emph{illustrative identity bridge}
\[
 T_C(c_C;\zeta_C^{\mathrm{id}})=c_C.
\]
This analyst-imposed correspondence is not an empirical transport result. On the log scale,
\begin{equation}
 z_C=\log(1.09)=0.0862,
 \qquad
 s_C\simeq\frac{\log(1.10)-\log(1.08)}{2\times1.96}=0.00468.
 \label{eq:gomesanchor}
\end{equation}
The reconstruction treats the published interval as approximately symmetric on the log scale. Under the identity bridge,
\[
 z_C=T_C(c_C;\zeta_C^{\mathrm{id}})+\Delta_C+\varepsilon_C
     =c_C+\Delta_C+\varepsilon_C.
\]
The residual $\Delta_C$ is a stress parameter conditional on accepting $\zeta_C^{\mathrm{id}}$. Bounding $|\Delta_C|\le\delta$ gives
\begin{equation}
 \ell_C(\delta)=\max\{0,z_C-1.96s_C-\delta\},
 \qquad
 u_C(\delta)=z_C+1.96s_C+\delta.
 \label{eq:gomesbound}
\end{equation}
For the vehicle/geographic actuarial target $L_{\mathrm{base}}=1.220153$, the externally restricted set is
\begin{equation}
\begin{aligned}
 \mathcal C_{\mathrm{ext}}(L_{\mathrm{base}};\delta)
 =\bigl\{(c_E,c_C,c_B)\in\mathbb R_+^3:\;&
 c_E+c_C+c_B=L_{\mathrm{base}},\\
 &\ell_C(\delta)\le c_C\le u_C(\delta)\bigr\}.
\end{aligned}
 \label{eq:workedpolytope}
\end{equation}
Table~\ref{tab:transportworked} uses $\delta=0$, $\log(1.05)$ and $\log(1.10)$. The first row reproduces study uncertainty under the imposed identity bridge; the other rows allow multiplicative discrepancies of 1.05 and 1.10. None is an empirically calibrated transport bound. Calibration would require comparable target-scale estimates or an explicit transport model.

\begin{table}[H]
\centering
\small
\begin{tabular}{lccc}
\toprule
Transport sensitivity & $\delta$ & bound on $c_C$ & equivalent ratio range \\ 
\midrule
Study uncertainty only & 0 & $[0.077,\,0.095]$ & $[1.080,\,1.100]$ \\ 
Allow factor 1.05 & $\log(1.05)$ & $[0.028,\,0.144]$ & $[1.029,\,1.155]$ \\ 
Allow factor 1.10 & $\log(1.10)$ & $[0.000,\,0.191]$ & $[1.000,\,1.210]$ \\ 
\bottomrule
\end{tabular}

\caption{Worked transport-sensitivity bounds obtained from the environmental/vehicle indirect-path estimate of \citet{gomesfranco2020age}. The last two rows enlarge the study uncertainty by a user-specified log-scale discrepancy. They are sensitivity scenarios, not confidence intervals for transportability.}
\label{tab:transportworked}
\end{table}

\begin{figure}[H]
\centering
\includegraphics[width=0.78\textwidth]{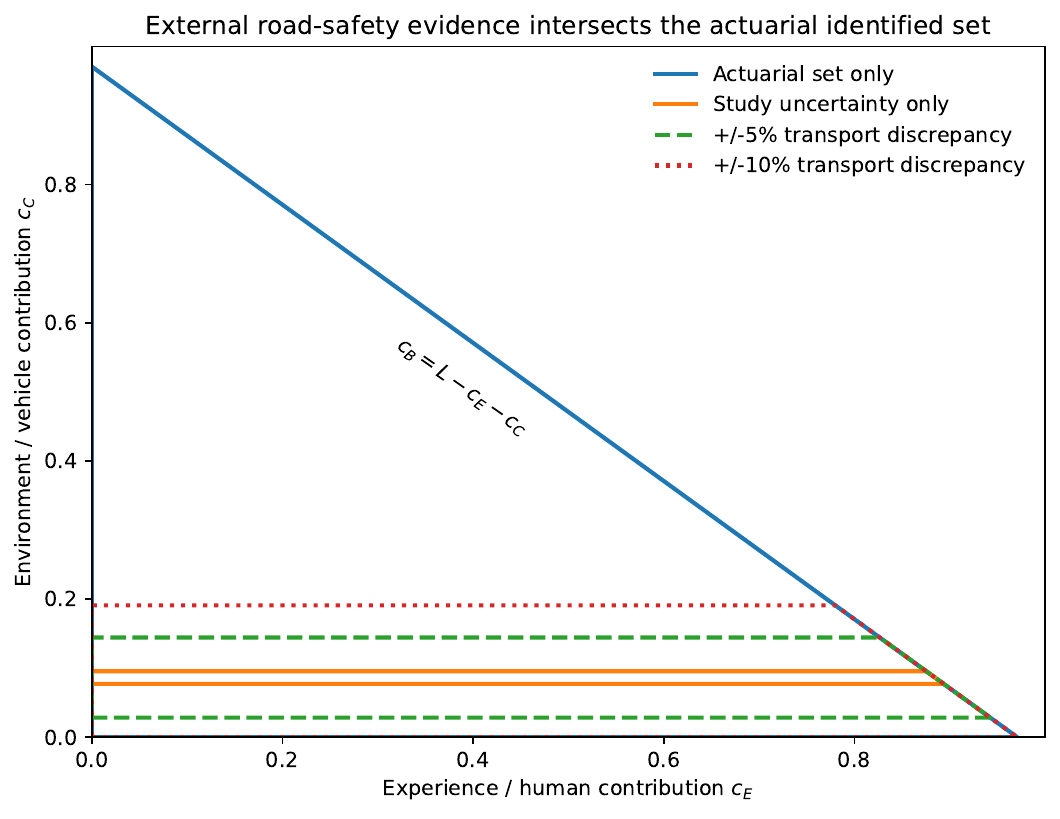}
\caption{Worked coarse compatibility region for the vehicle/geographic age relativity after adding external road safety evidence. The outer triangle is $\mathcal C_+(L_{\mathrm{base}})$ with $L_{\mathrm{base}}=\log(3.387706)=1.220153$; the nested polygons impose the environmental/vehicle indirect-path anchor from \citet{gomesfranco2020age} under increasingly permissive transport discrepancies. These polygons are conditional bookkeeping sets under an imposed sign restriction and bridge scenario; they are not graph-specific projections, posterior probability regions, or evidence that the Spanish mediation estimand transports to the French portfolio.}
\label{fig:externalcalibration}
\end{figure}

Figure~\ref{fig:externalcalibration} holds $L_{\mathrm{base}}$ fixed and restricts only $c_C$. Considerable width remains because experience, behaviour and residual mechanisms are not separated. The three-block display is not a projection of $\mathcal F_G(r)$ and cannot compare $G_0$ with $G_+$.

\subsection{What the quantitative examples identify}

The examples constrain different objects: an aggregate mileage elasticity, conditional claim-frequency contrasts, and one bookkeeping term under an imposed cross-study bridge. Sampling uncertainty in the French GLM does not resolve uncertainty about the crash-to-claim map, and a precise external estimate does not resolve observationally equivalent pathways. The remaining width records that unresolved structure; it is not a causal percentage.

\section{DECISION RELEVANCE AND DATA DESIGN}
\label{sec:validation}

A narrow interval for an annual rating contrast can coexist with substantial uncertainty about the mechanism that produced it. The practical importance of that ambiguity depends on the decision.

\subsection{When does mechanistic uncertainty matter for a decision?}

Let $a\in\mathcal A$ denote a prevention, monitoring, or data-collection action and $L(a;\psi)$ a loss under mechanism specification $\psi$. For compatibility set $\mathcal F$, define
\begin{equation}
 \mathcal A^{\star}(\mathcal F)
 =\bigcup_{\psi\in\mathcal F}\arg\min_{a\in\mathcal A}L(a;\psi).
 \label{eq:decisioninterface}
\end{equation}
We do not estimate $L$ or solve a policy problem. Equation~\eqref{eq:decisioninterface} only makes the decision link explicit. If the same action is preferred across compatible mechanisms, finer causal resolution may matter little for that decision. If preferred actions differ, measurements that separate those mechanisms may be useful once collection costs and measurement error are specified.

\subsection{A conceptual data-prioritization principle}

Sensitivity analysis can guide the collection of additional information \citep{frey2002sensitivity}. Here the corresponding question is simple: which measurements would reduce uncertainty in directions that could change an action? For motor risk, useful candidates include trip context, transient behavioural and conflict states, and linked crash--responsibility--claim records. Telematics can observe part of the middle layer, but it does not measure every latent mechanism.

This logic extends beyond insurance whenever the operational outcome is aggregated or administratively downstream from the physical event. Prediction and mechanism then answer different questions, and additional data are valuable only insofar as the distinction matters for the decision at hand.

\section{DISCUSSION}
\label{sec:discussion}

Motor insurance makes the resolution problem concrete. Annual models estimate claim risk; road-safety mechanisms operate within trips; reporting, coverage and responsibility assignment intervene between a crash and a recorded liability claim. A precise rating coefficient can therefore coexist with broad uncertainty about the process that generated it.

The age example illustrates this point without resolving it. The 18--20 versus 40--49 relativity is 3.388 in the vehicle/geographic specification and 1.235 when the model also conditions on the medium bonus--malus discretisation. The latter is a different conditional claim contrast, not a mediated age effect. The Gomes-Franco exercise is weaker still: it restricts one coarse bookkeeping term only after imposing a strong bridge from a Spanish culpability OR to the display scale. Its value is to show where the bridge assumption enters, not to validate transportability.

The limits are substantial. The DAG compresses within-trip dynamics into a time-ordered snapshot, and the evidence map is targeted, non-exhaustive, and adjudicated by one author. Road-safety studies use heterogeneous outcomes, while the insurance data observe claims rather than crashes. The crash-to-claim map is only partially specified. Finally, the three-block age geometry is a toy accounting model, not a projection of the nonparametric DAG. These restrictions are part of the analysis, not details to be hidden behind a point estimate.

The main implication is modest. Predictive and mechanism-specific risk statements require different evidence. Compatibility sets provide a way to keep that distinction visible and to identify which additional assumptions or measurements would be needed before a mechanism-specific interpretation is defensible.

\section{CONCLUSION}
\label{sec:conclusion}

Annual risk prediction and short-horizon risk mechanisms need not be resolved at the same level. In motor insurance, moving from a claim relativity to a statement about crash mechanisms requires assumptions about latent driving states, the aggregation of opportunities, the crash-to-claim process, and any external evidence used for calibration.

The proposed compatibility sets record those assumptions rather than replacing them with a single explanation. They are useful when the question is not only whether a model predicts risk, but what that prediction can support about mechanism. If different compatible mechanisms imply different actions, data closer to the risk-generating process become more valuable. If they imply the same action, the unresolved mechanism may matter less.

\section*{DATA AVAILABILITY}
The empirical illustration uses the pre-existing open \texttt{freMTPL2freq} portfolio from \texttt{CASdatasets}; no proprietary individual-level data were collected for this study. Code and computational outputs are available at \url{https://github.com/freakonometrics/causal_accident}; a rendered reproducibility appendix is available at \url{https://freakonometrics.github.io/causal_accident/}.

\clearpage
\appendix
\section*{Appendix overview}
The material below reproduces the Supporting Information prepared with the journal manuscript. It documents the evidence-extraction schema, the scope of the evidence-informed graph construction, methodological positioning, detailed evidence tables, and graph-sensitivity alternatives.
\addcontentsline{toc}{section}{Appendix overview}

\section{Study-level evidence-extraction schema}
\label{app:extraction}

Each row of the evidence database corresponds to an estimand-bearing study--edge relation, so a publication may appear more than once when it contributes distinct exposures or outcomes. The recorded fields are:
\begin{enumerate}[label=(\arabic*)]
\item bibliographic identifier and country;
\item study population and sampling frame;
\item design (naturalistic, cohort, case-control, responsibility analysis, before-after, quasi-experimental, or meta-analysis);
\item source node and target node;
\item temporal scale and temporal ordering;
\item outcome definition: crash, near-crash/conflict, responsibility, injury or severity;
\item effect measure, estimate, standard error or confidence interval;
\item adjustment set and potential over-adjustment variables;
\item measurement mode (objective, administrative, telematics, self-report);
\item whether selection is conditional on crash occurrence or severity;
\item whether a causal interpretation is supported by design and assumptions;
\item evidence-quality / risk-of-bias rating;
\item adjudication status (causal / proxy-only / uncertain / excluded), rationale, and decision to retain, merge, redirect or reject the edge in the integrated DAG.
\end{enumerate}

\section{Scope and traceability of the evidence-informed graph construction}
\label{app:search}

The evidence map supports construction and criticism of the DAG; it is not a systematic-review contribution and does not claim full compliance with a prospectively registered ESC-DAG review protocol. Four search families were used: (i) proximal crash mechanisms; (ii) driver experience and exposure; (iii) actuarial variables and intermediate mechanism measurements; and (iv) activity or mobility environment. Searches combined source-mechanism terms with occurrence, responsibility, conflict/near-crash or safety-critical outcomes and design terms appropriate to the question. The fixed search update used for this manuscript is 8 August 2026.

Traceability is defined at the level of each retained relation through the cited publication, sampled population, design, native outcome, effect measure, recorded threats to validity, adjudication rationale and quantitative-eligibility decision. The public computational archive accompanying this preprint reproduces the numerical illustrations; it is not presented as a public systematic-review database. The Appendix instead documents the extraction schema, adjudication rules, detailed evidence tables and graph-sensitivity alternatives used in the manuscript.

The 72 study--edge records are not treated as a random sample of the road-safety literature, and search completeness is not converted into a numerical uncertainty term. The tables below make retained relations and adjudication decisions inspectable, but they do not reconstruct a PRISMA-style search universe; the manuscript therefore makes no claim that retrieval completeness itself is reproducible. Q1 and Q2 studies can enter quantitative sensitivity analysis only after an estimand bridge is specified; Q3 records affect graph structure only; class P records belong to the observation layer. Because extraction and adjudication were conducted by one author, no inter-rater reliability statistic is available. The DAG should therefore be read as an evidence-informed structural model whose assumptions are open to re-adjudication, not as a definitive ontology of crash causation.

\section{Relation to adjacent methodological literatures}

The main text uses established tools selectively rather than presenting the compatibility construction as a replacement for them. Table~\ref{tab:positioning} summarizes the boundary.

\begin{table}[H]
\centering
\scriptsize
\begin{tabularx}{\textwidth}{@{}p{2.7cm}p{3.0cm}X@{}}
\toprule
Literature & Standard question & Role and boundary in this paper \\
\midrule
Partial identification & Which parameter values remain possible under maintained restrictions? & Motivates set-valued conclusions; the $\delta$-indexed bridge sets are sensitivity-feasible sets, not automatically sharp identified sets \cite{manski2003partial,tamer2010partial}. \\
Formal transportability & Can a causal estimand be recovered across populations from explicit invariance assumptions? & Provides the benchmark for what a justified transport formula would require; no such formula is claimed for the Spanish-to-French worked example \cite{bareinboim2013transport,pearlbareinboim2014external}. \\
Confounding sensitivity & How strong would unmeasured confounding need to be to alter an association? & Related in spirit but distinct: $\delta$ here stresses cross-study estimand mismatch rather than confounding strength \cite{cinelli2020sensitivity,vanderweele2017evalue}. \\
Causal mediation & Under what assumptions are direct and indirect effects identified? & Clarifies why a published mediation OR cannot be re-labelled as a claim-frequency path contribution without additional bridge assumptions \cite{imai2010mediation}. \\
Actuarial reporting and experience rating & How do accidents become reported claims, and how does past claim history enter prediction? & Gives substantive content to the observation layer and to bonus--malus as a history summary \cite{boucher2009accidents,boucher2022bonusmalus}. \\
Insurance fairness / causal structure & How can pricing information carry direct, indirect, or proxy content? & Adjacent evidence that predictive information sets need not coincide with structurally meaningful information; fairness is not the target estimand here \cite{lindholm2022discrimination}. \\
\bottomrule
\end{tabularx}
\caption{Methodological positioning. The paper combines these perspectives but does not claim to subsume their identification or decision-theoretic results.}
\label{tab:positioning}
\end{table}

\section{Detailed evidence-map tables}
\label{app:evidencetables}

The following tables preserve the detailed evidence map used to construct the graph. They are placed in the appendix because their role is traceability and adjudication rather than the main line of the quantitative argument.

\subsection{Evidence map for the structural core}

Table~\ref{tab:edges} lists the relations retained in, or kept adjacent to, the structural core. Broad rating labels are included only when they correspond to a more proximal mechanism.

\begin{landscape}
\small
\begin{longtable}{@{}p{2.4cm}p{2.3cm}p{2.6cm}p{1.5cm}p{7.2cm}@{}}
\caption{Evidence map for the causal core after edge adjudication and primary-study backtracking. ``Causal'' indicates that the relation is retained as a solid arrow in the integrated DAG; other statuses are kept outside the causal core or represented only indirectly.}\label{tab:edges}\\
\toprule
Source & Target & Evidence type & Status & Current interpretation \\
\midrule
\endfirsthead
\toprule
Source & Target & Evidence type & Status & Current interpretation \\
\midrule
\endhead
Exposure intensity & number of driving opportunities & exposure synthesis & causal & Annual distance is an imperfect but informative measure of opportunity count; crash counts rise less than proportionally with distance because context composition also changes \cite{elvik2023mileage,chipman1993exposure}. \\
Mean speed & crash occurrence / conflict & synthesis + matched telematics evidence & causal & A strong monotone speed--safety relationship is supported by the broader synthesis and by telematics case-control evidence linking speeding behaviour to crash risk \cite{elvik2019speed,winlaw2019telematics}. \\
Traffic state / speed variation & crash occurrence / conflict & systematic review and meta-analysis & causal & Short-term traffic characteristics precede and predict crash occurrence, supporting traffic as a contextual parent of conflict \cite{rosandel2015traffic}. \\
Adverse weather & crash occurrence / conflict & meta-analysis + exposure-adjusted road study & causal & Precipitation and road-surface conditions alter the road environment and crash rate, with important road-type interactions \cite{andrey1993rain,eisenberg2004precip,keay2006rain,blackmote2015winter,qiu2008weather,malin2019weather}. \\
Sleepiness / fatigue & crash occurrence & case-control, case-crossover and meta-analysis & causal & Acute case-control and case-crossover designs operate at the correct temporal scale, while later meta-analysis provides pooled support \cite{connor2002sleepiness,valent2010sleep,nabi2006sleepy,tefft2018sleep,dingus2016naturalistic,moradi2019sleepiness}. \\
Visual/manual distraction & safety-critical event & naturalistic systematic review and meta-analysis & causal & Short-duration distraction affects attention and conflict at the appropriate time scale \cite{klauer2014distracted,dingus2016naturalistic,simmons2016distraction}. \\
Passenger configuration & young-driver crash risk & systematic review & uncertain & Passenger effects are reproducible in young drivers but combine distraction, social influence and exposure selection; the graph retains the passenger-to-distraction pathway with lower confidence \cite{chen2000passengers,ouimet2015passengers}. \\
Night / passenger restrictions & fatal crash occurrence & quasi-experimental policy evaluation & causal & Graduated-licensing restrictions support nighttime/passenger exposure as modifiable components of novice-driver risk \cite{fell2011gdl}. \\
Errors / violations & crash involvement & meta-analysis & uncertain & The relation is useful mechanistically, but self-report, common-method bias and heterogeneous adjustment limit causal strength \cite{dewinter2010dbq}. \\
Latent risky-driving state & crash / near-crash & prospective hidden-state model & uncertain & Hidden-state models support persistent/evolving heterogeneity, but the state is model-defined rather than directly manipulable \cite{jackson2015hmm}. \\
Daily activity / travel pattern & route / time / exposure composition & activity-based model with trip diaries & causal & Activity and travel choices generate trips and therefore shape route, timing and exposure composition; the crash association is not used as a direct activity--crash coefficient \cite{elias2010activity}. \\
Vehicle telematics state & insurance loss & hidden Markov model & proxy-only & Important for validation and aggregation, but a learned telematics state is a measurement construct rather than a causal primitive \cite{jiang2024hmm}. \\
Vehicle safety technology & rear-end crash & replicated field-effectiveness studies & causal & Collision-avoidance technology is a proximal vehicle mechanism and is kept separate from broad vehicle class; comparable reductions recur across vehicle classes \cite{fildes2015aeb,cicchino2017aeb,cicchino2023aeb}. \\
\bottomrule
\end{longtable}
\end{landscape}

\subsection{Primary-study anchors and design heterogeneity}

Table~\ref{tab:primaryanchors} shows the value of retaining primary estimands. Chen et al.'s passenger dose-response concerns fatal-driver crashes per trip, while Tefft's outcome is culpable involvement conditional on a crash \cite{chen2000passengers,tefft2018sleep}. Both are informative, but neither is identical to the target crash-occurrence probability.

Several primary studies also sharpen the shape of local relations. In an urban matched case-control study, Kloeden et al. found that casualty-crash involvement approximately doubled for each 5~km/h increase above 60~km/h \cite{kloeden1997speed}. Freeway studies show that mean speed, speed variation and volume interact differently across traffic regimes \cite{golob2004traffic,xu2012trafficstate}. Observational and naturalistic studies link teenage peer passengers to speeding, short headways and other risky behaviours \cite{simonsmorton2005passengers,goodwin2012passengers}. In the DRIVE cohort, sleeping six hours or less per night was associated with an adjusted crash risk ratio of 1.21 (95\% CI 1.04--1.41) among young drivers \cite{martiniuk2013sleep}. Each estimate is retained on its native design and outcome scale.

\begin{table}[H]
\centering
\scriptsize
\begin{tabularx}{\textwidth}{@{}p{2.15cm}p{0.65cm}p{2.35cm}p{2.35cm}X@{}}
\toprule
Study & Class & Contrast & Estimate & Quantitative-use role / warning \\
\midrule
Nabi et al. \cite{nabi2006sleepy} & Q2 & sleepy driving a few times/year vs never & RR $1.50$ [1.20, 2.00] & Prospective long-horizon propensity; separate from acute state evidence. \\
Nabi et al. \cite{nabi2006sleepy} & Q2 & sleepy driving monthly or more vs never & RR $2.90$ [1.30, 6.30] & Dose-response, but self-reported exposure/outcome and occupational cohort. \\
Tefft \cite{tefft2018sleep} & Q1 & 6, 5, 4, $<4$ h sleep vs 7--9 h & OR $1.30$, $1.90$, $2.90$, $15.10$ & Acute dose-response close to responsibility target; conditions on crash involvement. \\
Klauer et al. \cite{klauer2014distracted} & Q2 & novice dialing / texting & OR $8.32$ [2.83, 24.42] / $3.87$ [1.62, 9.25] & Task-specific naturalistic crash-or-near-crash effects; map to conflict branch. \\
Dingus et al. \cite{dingus2016naturalistic} & Q2 & observed distraction / drowsiness / speeding & OR $2.0$ / $3.4$ / $12.8$ & Actual-crash naturalistic evidence; keep mechanism categories distinct. \\
Andrey--Yagar \cite{andrey1993rain} & Q2 & rain vs normal conditions & crash risk about $1.70$ & Aggregate event contrast; traffic adaptation/exposure requires sensitivity. \\
Black--Mote \cite{blackmote2015winter} & Q2 & winter precipitation event & collisions about $1.19$ & Matched city-event evidence; climate/adaptation transportability. \\
Fildes et al. \cite{fildes2015aeb} & Q1 & low-speed AEB equipped vs comparison & RR about $0.62$ & Direct branch-specific technology anchor for rear-end crashes. \\
Chen et al. \cite{chen2000passengers} & Q2 & 1 / 2 / 3+ passengers, age 16 & RR $1.39$ / $1.86$ / $2.82$ & Strong dose-response but fatal-driver-crash incidence mixes occurrence and severity. \\
\bottomrule
\end{tabularx}
\caption{Illustrative primary-study estimates recovered by backtracking from synthesis-level evidence. These estimates are not pooled across rows; they enter design- and outcome-specific evidence strata.}
\label{tab:primaryanchors}
\end{table}

Fatigue shows why a single pooled coefficient would be misleading. Acute case-control evidence reports ORs of 8.2 for self-reported sleepiness and 2.7 for five hours of sleep or less in the preceding 24 hours \cite{connor2002sleepiness}; the broader meta-analysis reports OR 1.34 for drowsy driving \cite{moradi2019sleepiness}; naturalistic crash data give OR 3.4 for observed drowsiness \cite{dingus2016naturalistic}; and the DRIVE cohort reports RR 1.21 for habitual short sleep \cite{martiniuk2013sleep}. Tefft adds a steep acute dose-response in a culpability design that conditions on crash involvement \cite{tefft2018sleep}. These estimates concern different exposures, populations, sampling schemes and outcomes. The main text therefore allows numerical reuse only after an explicit bridge specification $T_{d,o,\mathsf{pop}}(\theta;\zeta)$. When no scientifically defensible bridge is available, the estimate remains structural evidence and is not pooled or converted into a target-scale coefficient.

\subsection{Evidence for the actuarial bridge}
\label{sec:bridge-evidence}

For the observation layer, the relevant question is whether conventional rating factors carry information about the proximal variables in the crash DAG. The evidence is cleaner for two latent bridge objects than for a collection of direct links from coarse labels to individual contexts. We use $A_i$ for a stable activity/use profile and $G_i$ for the mobility environment associated with residence and habitual activity space. Declared use is modelled as a noisy measurement of $A_i$, and residential territory as a noisy measurement of $G_i$.

\begin{landscape}
\small
\begin{longtable}{@{}p{2.5cm}p{2.8cm}p{2.4cm}p{1.5cm}p{6.7cm}@{}}
\caption{Evidence map for the actuarial bridge. ``Proxy-only'' links belong to the observation layer and are never rendered as solid causal arrows.}\label{tab:bridge}\\
\toprule
Rating information & Proximal object & Design & Status & Interpretation for the observation layer \\
\midrule
\endfirsthead
\toprule
Rating information & Proximal object & Design & Status & Interpretation for the observation layer \\
\midrule
\endhead
Age / licence tenure & accumulated experience & naturalistic cohort / longitudinal & proxy-only & Licence tenure and practice directly order accumulated experience; age alone does not measure the same quantity when tenure is observed \cite{mccartt2003experience,gulliver2013learner,ehsani2020learner}. \\
Age & time-of-day exposure & exposure-adjusted crash study & proxy-only & Age-specific risk varies strongly by time of day, supporting age as information about exposure composition rather than a single direct age--crash mechanism \cite{regev2018agegender}. \\
Sex & annual mileage / exposure & national travel and crash data & proxy-only & Part of the raw sex difference in crash involvement is explained by different average annual mileage \cite{massie1997gender,chipman1993exposure}. \\
Sex & distance / driving habits & PAYD telematics & proxy-only & Detailed usage information can substantially reduce the incremental predictive role of sex \cite{ayuso2016gender}. \\
Age, sex & speeding distribution & insurance telematics & proxy-only & Speeding distributions vary with conventional covariates and observed context, providing a direct empirical bridge to a proximal mechanism \cite{guillen2021speeding}. \\
Young-driver sex & speeding / nighttime patterns & PAYD telematics with at-fault outcome & proxy-only & GPS data show distinct pattern distributions and distinct links from those patterns to distance before an at-fault crash \cite{ayuso2016gps}. \\
Annual mileage & exposure composition & PAYD / GPS exposure models & proxy-only & Distance does not exhaust exposure information; road class, timing and route/activity pattern add risk information \cite{paefgen2014exposure,ayuso2014firstaccident,ayuso2019ratemaking}. \\
Declared use & activity / travel profile $A_i$ & trip-diary activity model & proxy-only & Trip purpose and daily activity generate different route, time and distance patterns; the insurance use label is treated here as a coarse measurement of that activity profile \cite{elias2010activity}. \\
Business / work use & scheduling, fatigue and exposure profile & systematic reviews & uncertain & Work-related driving studies repeatedly identify duration, sleepiness, occupational stress and organisational conditions, but journey-purpose measurement is inconsistent and populations are heterogeneous \cite{robb2008work,newnam2011work,newnam2022systems}. \\
Residential territory & road / speed / environment profile $G_i$ & exposure surveys / crash-residence studies & proxy-only & Region and residence are associated with the environments in which people drive; differences in typical speed and environment explain part of regional crash-rate contrasts \cite{chipman1993exposure,blatt1998residence,lee2014residence}. \\
Territory / location & contextual telematics & GPS trajectories + traffic data & proxy-only & Location, traffic, peak-time travel and route context add information beyond driver demographics, supporting $G_i$ as a measurable intermediate layer \cite{ma2018context,guillen2024weekly}. \\
Territory & annual insurance risk & spatial motor-insurance model & proxy-only & Spatial rating models confirm that residence/location predicts annual claim frequency, but do not identify which road safety mechanism explains the spatial signal \cite{shi2017territorial}. \\
Vehicle power & operating speed & field speed observations & uncertain & Higher power is associated with higher operating speeds, but driver--vehicle selection prevents a simple one-edge causal interpretation \cite{mccartt2017power}. \\
Vehicle age / weight / registration year & KSI / injury crash outcome & population injury-crash models & uncertain & Høye models killed-or-seriously-injured outcomes; the study supports physical safety relevance but does not separate crash occurrence from injury severity, so it is not used as a clean occurrence edge \cite{hoye2019vehicle}. \\
Vehicle characteristics & harsh braking & naturalistic telematics & uncertain & Vehicle characteristics predict measured braking in a small cohort, illustrating both physical mechanisms and selection \cite{boylan2025vehicle}. \\
High-performance vehicle class & crash involvement & population restriction study & uncertain & The contrast mixes vehicle capability and driver selection and is therefore not retained as a clean solid edge \cite{keall2013performance}. \\
Traditional rating factors & telematics behaviour & insurance cohorts & proxy-only & Telematics adds information to conventional variables and can make demographic predictors redundant, directly supporting the observation-layer interpretation \cite{guillen2019rates,henckaerts2022dynamic,huang2019classification}. \\
Past behavioural / claims signals & persistent state $H_i$ & longitudinal credibility / telematics & proxy-only & Repeated history updates the distribution of persistent heterogeneity rather than physically causing the next crash \cite{denuit2019credibility}. \\
\bottomrule
\end{longtable}
\end{landscape}

The retained evidence does not justify direct declared-use or territory arrows to crash. Activity-based models link daily activities and travel choices to route, timing and exposure \cite{elias2010activity}; work-related-driving reviews point to scheduling, fatigue and organisational mechanisms but remain heterogeneous \cite{robb2008work,newnam2022systems}. Residence and region predict the road environments in which driving occurs, and regional contrasts change when time, distance, speed and environment are handled \cite{chipman1993exposure,blatt1998residence,lee2014residence}. These findings are consistent with using $A_i$ and $G_i$ as intermediate measurement targets under the stated observation model.

\subsection{Reduced-form shortcuts: excluded or uncertain}

Table~\ref{tab:excluded} records reduced-form shortcuts that are not imposed in the structural graph, even though they may appear in informal tariff explanations.

\begin{table}[H]
\centering
\small
\begin{tabularx}{\textwidth}{@{}p{3.0cm}p{1.7cm}X@{}}
\toprule
Candidate shortcut & Status & Reason \\
\midrule
Past claims $\to$ future crash & excluded & Reversed measurement logic: persistent heterogeneity generates both past and future outcomes. \\
Sex $\to$ crash & excluded & Predictive information is routed through exposure/behaviour distributions; no single proximal mechanism is asserted. \\
Territory $\to$ crash & excluded & Territory is a coarse measurement of mobility environment $G_i$, not a physical crash mechanism. \\
Declared use $\to$ crash & excluded & Use is treated as a measurement of activity profile $A_i$, which changes route, timing, exposure and work constraints. \\
Vehicle model $\to$ crash & excluded & The coarse label combines performance, safety technology, mass/structure and driver selection; these components must be separated. \\
Age $\to$ crash & uncertain & A residual direct path may summarize unmodelled physiological or cohort mechanisms, but the main text prioritizes mediated paths through experience, exposure and behaviour. \\
\bottomrule
\end{tabularx}
\caption{Reduced-form arrows rejected or left uncertain in the integrated DAG.}
\label{tab:excluded}
\end{table}

\subsection{Feedback extension: insurance can also change behaviour}
\label{sec:feedback}

The observation-layer interpretation is most natural for conventional annual pricing, where the tariff records information about risk but does not directly intervene on driving. Usage-based contracts may break that separation. Field experiments show that financial incentives can change speed choice, and randomized telematics feedback or incentive interventions can change measured driving behaviour \cite{bolderdijk2011payd,stevenson2021feedback}. A dynamic insurance DAG may therefore contain the feedback path
\[
\text{contract design / feedback}_{t}
\longrightarrow
\text{behaviour}_{t+1}
\longrightarrow
\text{future risk}.
\]
This behavioural-feedback channel is outside the baseline graph. Once pricing changes subsequent driving, contract design belongs in the causal system and can no longer be treated as a passive observation.

\section{Formal graph specification and sensitivity set}
\label{app:graphspec}

A machine-readable adjacency representation was used to check the displayed relations by source, target, layer and status. The solid-edge core $G_0$ is used for the baseline compatibility analysis; densely dotted relations are candidate additions. Table~\ref{tab:graphalternatives} records the main alternatives and the part of the decomposition they affect.

\begin{table}[H]
\centering
\small
\begin{tabular}{p{0.24\textwidth}p{0.15\textwidth}p{0.51\textwidth}}
\toprule
Candidate relation & Status & Consequence if admitted \\
\midrule
Passengers $\to$ distraction & uncertain & Adds a context-to-transient-state pathway; can reallocate part of an age/context contrast toward distraction without changing the annual aggregation identity. \\
Vehicle performance $\to$ speed & uncertain & Adds a vehicle-to-behaviour route; can move compatible contribution from residual behaviour to the vehicle/performance branch. \\
Errors $\to$ conflict & uncertain & Enlarges the behaviour-to-conflict mapping and therefore weakens restrictions that rely on conflict being driven only by the retained proximal parents. \\
Violations $\to$ conflict & uncertain & Same structural role as the errors edge, with a distinct behavioural construct and evidence base. \\
Residual age $\to$ crash & uncertain shortcut, not drawn in $G_0$ & Introduces an additional residual age block. It does not identify an intervention on age; it enlarges the family of decompositions compatible with the annual age relativity. \\
Persistent state $H_i\to S_{it}$ & structural sensitivity, not drawn in $G_0$ & Allows stable heterogeneity to affect transient state directly rather than only through behaviour. Admitting it relaxes the block-level restriction $S_{it}\perp H_i\mid(R_i,C_{it})$. \\
Experience $R_i\to C_{it}$ & structural sensitivity, not drawn in $G_0$ & Allows accumulated experience to alter context selection beyond the activity/use and mobility-environment profiles. Admitting it relaxes $C_{it}\perp R_i\mid(A_i,G_i)$. \\
Direct state/context $\to Y_{it}$ & proximal-state sensitivity, not drawn in $G_0$ & Relaxes the strong baseline assumption that $K_{it}$ is sufficient for crash occurrence given the displayed state. Selected direct parents may include context, transient state, behaviour, or vehicle safety. \\
\bottomrule
\end{tabular}
\caption{Graph-structure sensitivity for the principal relations not imposed in the solid-edge core. Within the nested nonparametric graph family defined in the main text, adding these arrows can only preserve or enlarge the compatible mechanism set for a fixed annual contrast.}
\label{tab:graphalternatives}
\end{table}

At the block level, the core parent sets used in the main text are $\mathrm{pa}(C)=\{A,G\}$, $\mathrm{pa}(S)=\{R,C\}$, $\mathrm{pa}(B)=\{R,H,C,S\}$, $\mathrm{pa}(K)=\{C,B,V^{\mathrm{safe}}\}$ and $\mathrm{pa}(Y)=\{K\}$, together with $M\mid(E,A,G)$. The sensitivity rows $H\to S$, $R\to C$, and direct parents of $Y$ are listed precisely because their absence is otherwise easy to read as an unexamined substantive claim. The finer adjacency file is authoritative for component-level edges such as night $\to$ fatigue or traffic $\to$ speed. This distinction prevents the compact block factorization from being mistaken for a claim that every member of a block has every block-level parent.

\bibliographystyle{apalike}
\bibliography{references}
\end{document}